# *dataRLsec*: Safety, Security, and Reliability With Robust Offline Reinforcement Learning for DPAs

Shriram KS Pandian, Dept. of Cybersecurity, Rochester Institute of Technology, Rochester, New York, USA 14623
Naresh Kshetri, Department of Cybersecurity, Rochester Institute of Technology, Rochester, New York, USA 14623

**Abstract** - Data poisoning attacks (DPAs) are becoming popular as artificial intelligence (AI) algorithms, machine learning (ML) algorithms, and deep learning (DL) algorithms in this artificial intelligence (AI) era. Hackers and penetration testers are excessively injecting malicious contents in the training data (and in testing data too) that leads to false results that are very hard to inspect and predict. We have analyzed several recent technologies used (from deep reinforcement learning to federated learning) for the DPAs and their safety, security, & countermeasures. The problem setup along with the problem estimation is shown in the MuJoCo environment with performance of HalfCheetah before the dataset is poisoned and after the dataset is poisoned. We have analyzed several risks associated with the DPAs and falsification in medical data from popular poisoning data attacks to some popular data defenses. We have proposed robust offline reinforcement learning (Offline RL) for the safety and reliability with weighted hash verification along with density-ratio weighted behavioral cloning (DWBC) algorithm. The four stages of the proposed algorithm (as the Stage 0, the Stage 1, the Stage 2, and the Stage 3) are described with respect to offline RL, safety, and security for DPAs. The conclusion and future scope are provided with the intent to combine DWBC with other data defense strategies to counter and protect future contamination cyberattacks.

## 1. Introduction

We are in the cyber world today (as cyber is one popular domain after land, sea, air, water) and also in the world of cyberattacks every second, data poisoning attacks (DPAs) are one of those popular cyberattacks. The security of data and our dataset is a must via several techniques and one of them is security via Machine Learning (ML) [1]. Data poisoning attacks and data poisoning threats are of more significance today as we are dealing with the web-based dataset. The concept of online poisoning with several data poisoning strategies are applied by attackers (and penetration testers) to manipulate the training data.

Whether it is a five layer network architecture model or the seven layer model, we are surrounded by protocols and the new protocols are growing every day. Privacy protocols for the data poisoning attacks (DPAs) is the one called Local Differential Protocol (LDP) for data analytics in privacy-preserving data [2]. In LDP protocol, hackers can send crafted data to the data collector via easily injecting fake users. As an emerging data threat (as machine learning is widely deployed), attackers are manipulating training data to degrade performance [3]. Malicious outcomes are enabled as a part of successful cyber-attacks (primarily data poisoning attacks and issues), from IP theft, autonomous systems hacking, backdoors injection, label flipping and many more.

The need of cyber defenses for DPAs is must and we have tried to provide that via robust offline RL. Quickly assessing defenses on a given dataset (as ML systems are susceptible), MNIST-1-7 and Dogfish datasets are attack-resilient even under a simple defense [4]. As we know that the training data comes directly from the outside world, attackers can easily inject malicious data via user account creation. Training only a mimic model to imitate the behavior of the target model via clean samples can be one solution to the data poisoning attacks (DPAs) and several variants of DPAs [5]. Different realistic datasets for several types of poisoning attacks on several defense methods can be effective and efficient. Machine learning algorithms are always susceptible to security threats as the popularity of these algorithms are growing every day.

Our remaining work on the dataRLsec project is as below in the following sections. Section 2 is the background work where we have identified five separate works with comparative analysis for data poisoning attacks (DPAs). Section 3 of our study is the problem and estimation of the MuJoCo environment where the dataset is poisoned with HalfCheetah performance plotted. The risks associated with data poisoning attacks are presented and outlined in Section 4 of the study. The section 5 of our work (Offline RL, Security, and Safety) deals with a proposed algorithm and explanation in four stages. The last section, Section 6 of the study is all about Limitations, Conclusion, and Future Scope of the study.

## 2. Background Work

Authors conducted experiments and analysis on real life data sets with problem modeling against crowdsensing systems of partially observable data poisoning attack [6]. Individual own mobile devices (primarily sensor embedded devices) that collect various data for crowdsensing systems. For evaluating the trustworthiness of data providers, frameworks like TruthFinder can resolve data conflicts. After collecting data from workers, cloud services can run a TruthFinder implementing algorithm such as the truth discovery algorithms.

Data poisoning attacks (DPAs) are vulnerable for multi-user semantic communication (MUSC) and researchers have proposed an effective attack-defense game framework known as DPAD-MUSC [7]. During image transmission for MUSC, DPAD-MUSC is tailored to defend against DPAs. Proliferation of IoT devices and advancement of Deep Learning there are data volumes (unprecedented data) susceptible to DPAs. While maintaining a higher evasion rate, DPAD-MUSC simulation results can find optimal attack policies with greater accuracy drop.

Researchers proposed a novel system-aware optimization method to derive poisoned data gradients and attain federated machine learning attack strategies [8]. Bilevel program is the problem of calculating optimal poisoning attacks adaptive to target and source attacking node(s) selection in an arbitrary way. Security and reliability of machine learning has always been a concern as it is used for IoT devices, natural gas price prediction, lesions segmentation, spam filtering and many others.

A deep reinforcement learning (RL) based framework for DPAs (data poisoning attacks) is proposed by authors as DRLAttack [9]. The proposed framework, DRLAttack, for precise targeting of data poisoning can generate more potent and stealthy fake user interactions to dynamically tailor attack strategies to context change recommendations. The training phase (initial phase) of the machine learning (ML) life cycle is the phase where data poisoning attacks (DPAs) aim to subtly change the training data.

For client selection and enhancing model robustness by choosing reliable peers in peer to peer federated learning, a deep reinforcement method is proposed by authors [10]. Throughout training the model exhibits more stable accuracy trends and reduced noise levels under dynamic and adaptability conditions. The need of the central coordinator (that is typically required in federated learning algorithms) is eliminated due to the distributed scheme of the combined deep reinforcement learning - federated learning (DRL-FL) algorithm.

Table 1: Comparative analysis of the Technology used with several insights of the background work w.r.t. Data Poisoning Attacks (DPAs)

| Ref./Year | Technology Used | Insight 1 | Insight 2 | Insight 3 |
|---|---|---|---|---|
| [6]/2020 | Deep Reinforcement Learning | Deep RL for partially observable in crowdsensing systems that collect various data from sensors. | Crowdsensing systems are susceptible to data poisoning attacks. | TruthFinder can resolve data conflicts but limit the impact of dirty data. |
| [7]/2024 | Adversarial Reinforcement Learning | Data poisoning attacks defense for task oriented multiuser semantic communication (MUSC). | Current techniques against DPAs are for traditional networks and not for MUSC. | Effective attack-defense game framework proposed as DPAD-MUSC. |
| [8]/2022 | Federated Machine Learning | Enables resource constrained devices to establish knowledge shared models. | Provides privacy preservation, economic benefit, & keeping raw data. | Propose a novel system aware optimization method, attack on FL. |
| [9]/2025 | Deep Reinforcement Learning (DRL) | Collaborative filtering is widely used & susceptible to data poisoning attacks. | Malicious actors inject synthetic user interaction data to manipulate results. | Propose DRLAttack, a deep reinforcement based framework for data poisoning. |
| [10]/2025 | Federated Learning (FL) with Deep RL (DRL) | Peer-to-peer (P2P) federated learning remains a critical challenge due to lack of centralized oversight. | Propose a deep reinforcement learning (DRL) method for client selection in P2P FL. | Framing client selection as a Markov Decision Process (MDP), enables clients to identify. |

## 3. Problem & Estimation

For the problem and its estimation, an environment of actual set up is done. We have set up the MuJoCo (Multi-Joint dynamics with Contact) environment [11] for the HalfCheetah (and performance is calculated). Figure 1 and Figure 2 shows the actual setup before the dataset is poisoned and after the dataset is poisoned. The degraded performance shown in Figure 2 (after the Minari dataset [12] is poisoned) of the HalfCheetah agent within the simulation environment. There are several simulation environments, but we have gone with MuJoCo simulation environment because of its relevant model optimization, accuracy, and speed. Despite being a free and open source simulation environment, MuJoCo is not only a simulator but also has interactive 3D visualization and multi-threaded sampling to facilitate research and development.

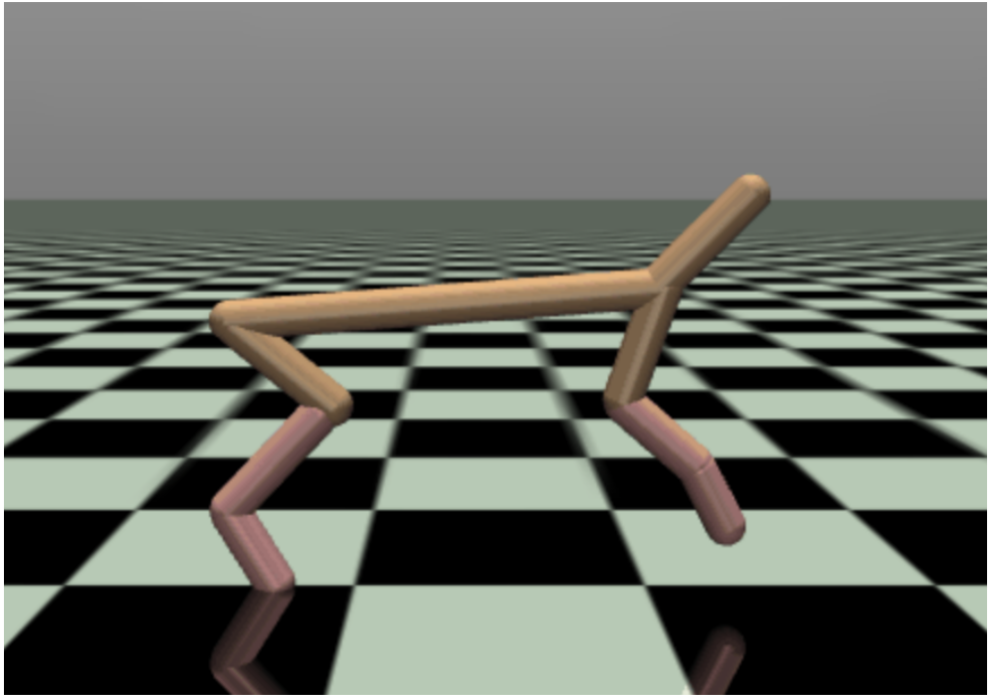

Figure 1: Depicting the MuJoCo environment with the actual setup where the performance of the HalfCheetah is calculated.

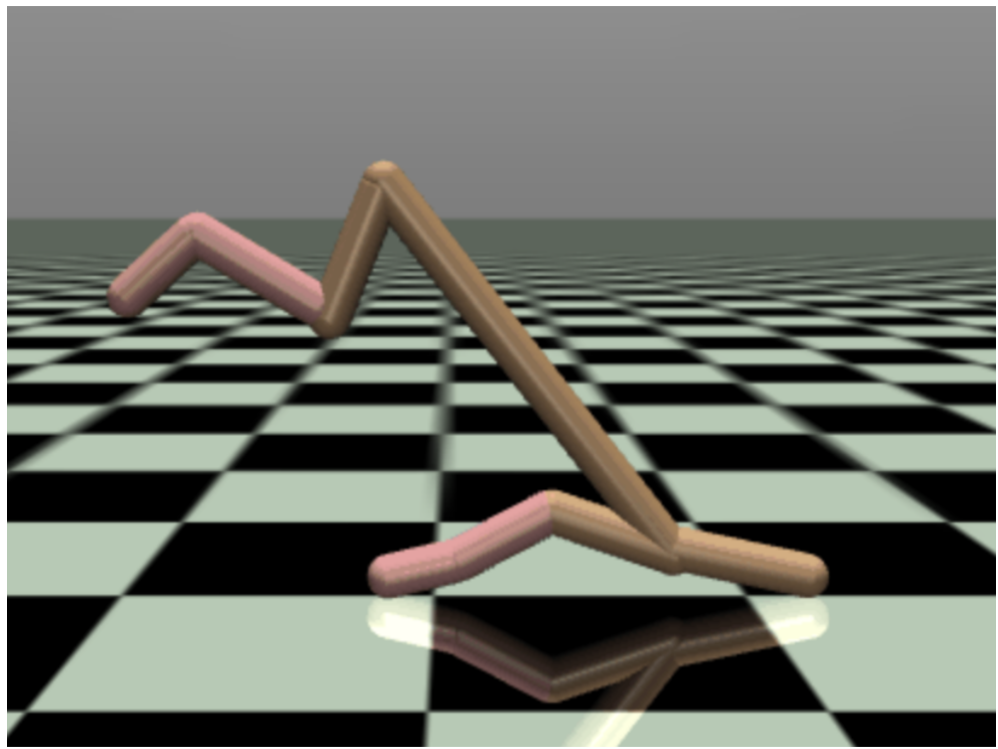

Figure 2: Leading to degraded performance of the HalfCheetah agent within the MuJoCo simulation environment, after the Minari dataset is poisoned.

## 4. Data Poisoning Attacks & Risks

As the use of machine learning algorithms has been widely used, security and safety has become a serious concern. Major summary of data poisoning attacks can be availability attack, targeted attack, and subpopulation attack for crowdsensing systems to sentiment analysis systems [13]. Deep learning and reinforcement learning are both used for the availability attack & targeted attack whereas Deep Learning is only actively used for the subpopulation attack. Some of the popular defenses can be Robustness enhancement, Data augmentation, Data sanitization, and Data Aggregation.

Insertion of malware into AI systems and machine learning algorithms have been increasing a lot. Sending false or misleading data inputs leads to several types (primarily six types) of data poisoning attacks (DPAs) as i. Stealth attacks, ii. Model Inversion attacks, iii. Training data poisoning, iv. Backdoor poisoning, v. Non-targeted attacks, and vi. Targeted attacks [14]. Unlike traditional cyberattacks, malicious actors manipulate training data in critical domains from healthcare to finance domains. DPAs have several consequences from amplifying vulnerabilities within affected systems to degrading model accuracy.

One popular dataset, The Pile for LLM development, is vulnerable to data poisoning attacks (DPAs) [15]. Hackers are using massive volumes of medical data to falsify medical knowledge as we know the concepts of GIGO in computer science (Garbage In Garbage Out). Some automated algorithms (say, quality control algorithms) can filter out undesirable data but may not account for syntactically hidden misinformation and false knowledge. Web scale datasets (Common Crawl, The Pile, Others) are always at risk of vulnerable pre-training data and AI models are easily compromised.

Some clean-label attacks can be added by attackers to enhance stealthies and avoid the detection in machine learning algorithms [16]. The boundary of the poisoned classifier is skewed by several means such as flipping the labels of training samples and mounting DPAs by malicious users. Detection is more challenging for human expert and ML models as the results are sample specific and invisible triggers. Open research challenges still continue in the area of DPAs as the fairness and reliability of a model that is composed of unfair inputs and undermine the trustworthiness dimensions of ML algorithms.

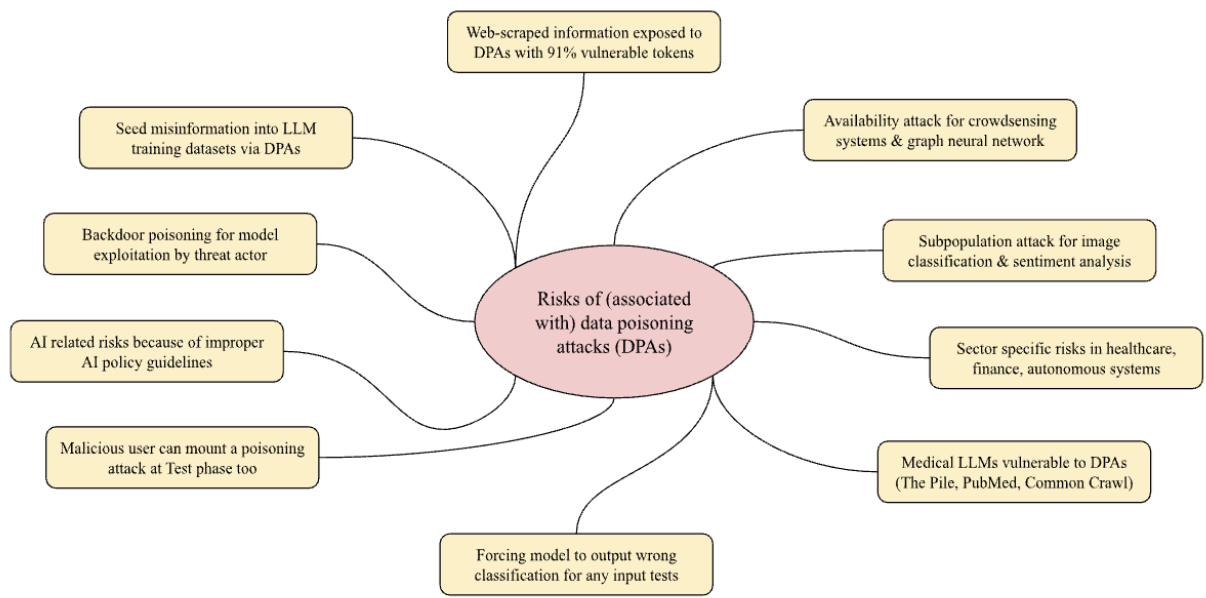

Figure 3: Summary of risks associated with data poisoning attacks (DPAs) [13] - [16]

## 5. Offline RL, Security, & Safety

As proposed, our offline reinforcement learning & safety, for data poisoning attacks (DPAs) is dataRLsec with behavioral cloning via hash verification in four stages below for Algorithm 1: Density-Ratio Weighted Behavioral Cloning with Hash Verification. The combination of several emerging technologies (say, blockchain, AI, quantum security, metaverse, deepfake, machine learning, & others etc.) is a must today in several AI-based systems for improving the safety and security [17] [18] (including fraud detection, data security via AI [19] [20]) and for the defense / counter of several ongoing cyber-attacks like cross-site request forgery (XSRF) attacks & cross-site scripting (XSS) attacks [21].

Stage 0: Verify Reference Set Integrity - After we start the algorithm with data (D as contaminated dataset & $D_{ref}$ as reference set), hyperparameters ($\epsilon$, & C), files & params ($H_{ref}$ as hash file, Hash algorithm, Integrity check mode), we will initiate Stage 0. The first decision is decided here in Stage 0 as "Hash verification enabled?" If the "Hash verification enabled?" is True / Yes, we will go for a second decision as "Integrity file $H_{ref}$ exists?". If the "Hash verification enabled?" is False / No, we will "Compute hash for $D_{ref}$ and save to $H_{ref}$". If the "Integrity file $H_{ref}$ exists?" is True / Yes, we verify $D_{ref}$ against stored hash $H_{ref}$ and go for the third decision. If the "Integrity file $H_{ref}$ exists?" is False / No, we will "Compute hash for $D_{ref}$ and save to $H_{ref}$". The third decision is "Verification successful?", and if True we will "Abort Training" else we will "End Stage 0".

Stage 1: Train Discriminator - We start Stage 1 and initialize the discriminator ($d_\phi$) with random weights. We check if "Epoch 1 to $E_d$?" is True or False after that initialization. If True / Yes, we perform three sets of operations as 1. Sample balance batch $B_{ref} \sim D_{ref}$ (label 1), 2. Sample balance batch $B_{main} \sim D$ (label 0), and 3. Update $\phi$ via binary cross-entropy loss. If False / No, we direct to 3. Update $\phi$ via binary cross-entropy loss. We then End the Stage 1 before starting the Stage 2.

Stage 2: Compute Weights - We start Stage 2 and check "For each trajectory?". If Yes, we perform three operations as 1. Compute density ratio: $r_i$, 2. Compute weight: $w_i$, and 3. Freeze weights $\{w_i\}$. If the check is No, we direct to 3. Freeze weights $\{w_i\}$. We then end Stage 2 before starting Stage 3.

Stage 3: Train Policy - The last stage is Stage 3 which is started via policy initialization with random weights. We then check "Epoch 1 to $E_\pi$?" after policy initialization. If True / Yes, we perform three operations as 1. Sample batch $B \subset D$, 2. Compute weighted loss $L_\pi$ using weights $w_j$, and 3. Update $\theta$ with gradient descent. If False / No, we direct to 3. Update $\theta$ with gradient descent. We then End Stage 3 and provide output as Robust policy $\pi_\theta$ and then end the algorithm.

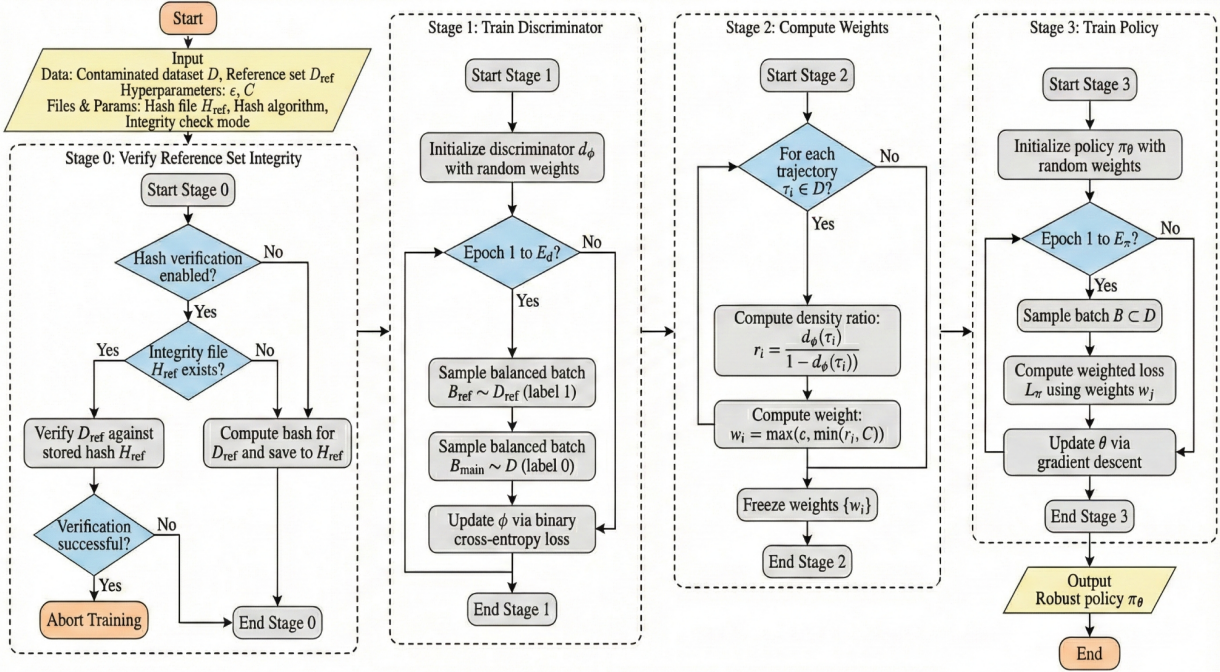

Algorithm 1: Density-Ratio Weighted Behavioral Cloning with Hash Verification

## 6. Limitations, Conclusion & Future Scope

Although Density-Ratio Weighted Behavioral Cloning (DWBC) demonstrates strong resistance to different kinds of poisoning attacks within offline reinforcement learning (Offline RL), there are still some flaws associated with this algorithm. Some of the limitations of the proposed algorithm of dataRLsec are pointed below.

a. To begin with, DWBC requires access to a reference dataset consisting of a small set of clean trajectories, which is not necessarily available within any domain. This is because the success rate of this algorithm is dependent on this dataset being reliable and representative.

b. Second, this algorithm is restricted to stationary contamination processes and static offline environments and is not capable of handling attacks themselves within dynamic settings.

c. Third, this algorithm is also likely to introduce a level of overhead due to the calculation involved by utilizing a discriminator to assign weights to trajectories.

d. Finally, although this algorithm is very resistant to different kinds of poisoning attacks, it could encounter difficulties when faced with very finely modeled attacks.

We have proposed a robust offline reinforcement learning (Offline RL) for the safety and reliability with respect to data poisoning attacks (DPAs). The Density-Ratio Weighted Behavioral Cloning (DWBC) algorithm with hash verification can train policy with random weights via reference set integrity, discriminator training, and weights computing. We have summarized and analyzed several risks associated within, from "web-scraped information exposed" to the "wrong output classification for any input tests". As shown in the problem setup of the MuJoCo environment and degraded performance of HalfCheetah, robust offline RL is much suitable for the countermeasures of the data poisoning attacks and its several variants attacks.

To generalize Density-Ratio Weighted Behavioral Cloning (DWBC's) capabilities, there are multiple lines of work to be pursued to enhance. Some of the improvements and future enhancements over the proposed dataRLsec are given below in a point-wise fashion.

a. To improve contamination detection and policy learning is one such line via the incorporation of reinforcement learning (RL) modules utilizing environment dynamics.

b. Adaptation to online domains dealing with continuously varying contamination shifts is another line to make Density-Ratio Weighted Behavioral Cloning (DWBC) amenable to. Improving density ratio estimation to make these approaches more efficient and scalable is also an interesting line to work on.

c. To better prove the effectiveness of Density-Ratio Weighted Behavioral Cloning (DWBC), further evaluations on more relevant domains like practical safety-critical controls can be done.

d. Finally, combining approaches like Density-Ratio Weighted Behavioral Cloning (DWBC) with other complementary strategies like adversarial training could lead to complete protection against contamination attacks.

Autonomous Systems (ICMLAS), Prawet, Thailand, 2025, pp. 793-798, doi: 10.1109/ICMLAS64557.2025.10968798.

[21] N. Kshetri, D. Kumar, J. Hutson, N. Kaur, and O. F. Osama, "AlgoXSSF: Detection and Analysis of Cross-Site Request Forgery (XSRF) and Cross-Site Scripting (XSS) Attacks via Machine Learning Algorithms," 2024 12th International Symposium on Digital Forensics and Security (ISDFS), San Antonio, TX, USA, 2024, pp. 1-8, doi: 10.1109/ISDFS60797.2024.10527278.
Page 10 of 10